\documentclass[pra,showpacs,superscriptaddress,twocolumn]{revtex4-1}
\usepackage{graphicx}
\usepackage{amsfonts}
\usepackage{amssymb}
\usepackage{amsmath}
\usepackage{epsfig}

\begin{document}

\title{Hybrid exceptional point created from type III Dirac point}
\author{L. Jin}
\email{jinliang@nankai.edu.cn}
\affiliation{School of Physics, Nankai University, Tianjin 300071, China}
\author{H. C. Wu}
\affiliation{School of Physics, Nankai University, Tianjin 300071, China}
\author{Bo-Bo Wei}
\email{weibobo@cuhk.edu.cn}
\affiliation{School of Science and Engineering, The Chinese University of Hong Kong,
Shenzhen, Shenzhen 518172, China}
\affiliation{Center for Quantum Computing, Peng Cheng Laboratory, Shenzhen 518055, China}
\author{Z. Song}
\affiliation{School of Physics, Nankai University, Tianjin 300071, China}

\begin{abstract}
Degeneracy points and exceptional points embedded in the energy band are
distinct by their topological features. We report hybrid exceptional point
formed through merging two ordinary exceptional points with opposite
chiralities that created from the type III Dirac points emerging from a flat
band. The hybrid exceptional point is induced by the destructive
interference at the proper match between the non-Hermiticity and the
synthetic magnetic flux. The degeneracy points and different types of
exceptional points are distinguishable by their topological features of
global geometric phase accompanied with the scaling exponent of phase
rigidity. Our findings not only pave the way of creating, moving, and
merging EPs but also shed light on the future investigations of
non-Hermitian topological phases.
\end{abstract}

\maketitle

\section{Introduction}

Exceptional points (EPs) are non-Hermitian degeneracies~\cite%
{KatoBook,Bender,NMBook,Miri}, at which the system Hamiltonian is defective
and eigenstates coalesce~\cite{Berry04}. Parity-time ($\mathcal{PT}$)
symmetric phase transition and many intriguing dynamical phenomena occur at
the EPs~\cite%
{Klaiman,Makris,AGuo,PengNP,Chang,Regensburger,LFeng,Alu,CPA,LYangNM,PXue,SLonghi,LFeng17,PTRev,ChenYF}%
. The properties of non-Hermitian system dramatically change in the vicinity
of EPs and are valuable for optical sensing \cite%
{Wiersig,NMNJP,ZPL,EP2Sensing,EP3Sensing}. The frequency sensing is enhanced
because that the responses of energy levels to the detuning perturbation are
square root near a two-state coalescence (EP2) and cubic root near a
three-state coalescence (EP3) in non-Hermitian systems, being more efficient
than the linear response near a diabolic point (DP) in Hermitian systems~%
\cite{EP2Sensing,EP3Sensing}.

The EPs possess distinct topology from DPs~\cite{Seyranian}. In a two-level
non-Hermitian system, the energy levels interchange after encircling the EP
for one circle in the parameter space; the interchanged energy levels
restore their original values after two circles of encircling and accumulate
a geometric phase $\pm \pi $; the sign of the geometric phase
depends on the circling direction and the chirality of the EP is defined by the accumulated geometric phase under the counterclockwise encircling~\cite%
{DembowskiPRL,Dembowski,Uzdin,Berry,Heiss12,SYL12,Milburn,TELee,KDingPRX,LJinPRA18}%
. Dynamical encircling of EPs realizes state switch and nonreciprocal
topological energy transfer~\cite{Zhen,Doppler,Xu,DNCPRL2017}; the dynamics
depends on the starting/end point~\cite{CTChanPRX18} and the homotopy of the
encircling loop \cite{GanainyNC18}.

The EPs are different by their ways of coalescence as well as their
topological properties~\cite{Heiss2008}. In one aspect, the EPs are distinct
by their orders: If more than two energy levels coalesce at the EP, the EP
is called a high order EP \cite{Graefe08,ZXZ}, where the excitation
intensity presents the polynomial increase \cite{QZhong}. In another aspect,
even the EPs with identical order may dramatically differ from each other on
the topological aspect. For example, when encircling a high order EP of
three-state coalescence (EP3) in the energy band of a square root type
Riemann surface in the parameter space, two energy levels flip after
encircling one circle; if an EP3 is in a cubic root type Riemann surface,
when it is encircled, three circles are needed for the energy levels to
restore their original values. The geometric phases associated with the two
EP3s are different, which reflect the different topological features of such
two types of EP3s~\cite{Graefe}. An interesting question naturally arises:
whether EP2s have different topologies? If so, how to characterize their
topological properties and distinguish them?

\begin{figure}[b]
\includegraphics[bb=0 0 485 75, width=8.8 cm, clip]{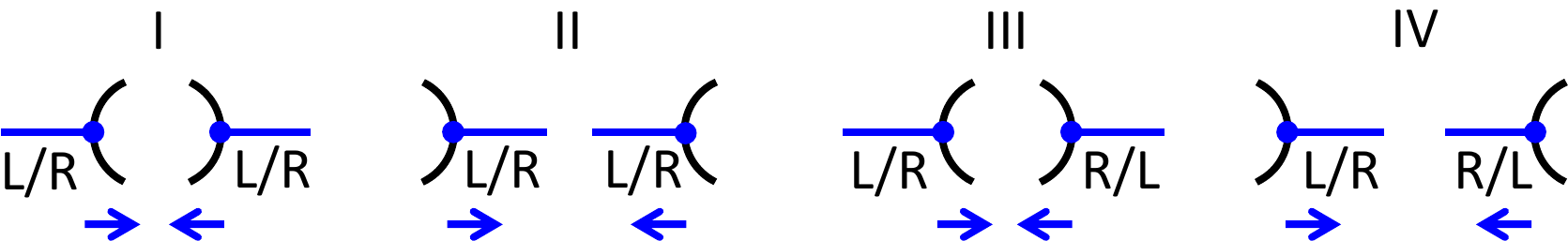}
\caption{Schematic of four types of EP2 merging indicated by the blue arrows.  The blue dots indicate the
left/right chiral EP2s embedded in the branched energy bands; the blue lines indicate the Fermi arcs. The chirality refers to the geometric phase associated with the EP2 when it is encircled in the parameter space.
The vertical axis indicates the real part of energy levels and the horizontal axis indicates a system parameter.}
\label{fig1}
\end{figure}

The manipulation of Dirac points in condensed matter physics is an
interesting and challenging task. The merging of Dirac points induces
topological phase transition and generates new types of Hermitian
degeneracy; for example, two Dirac points with opposite topological charges
can merge into a semi-Dirac point with linear and quadratic dispersions
along two orthogonal directions \cite{Montambaux,Esslinger,JMHou,SLin19}. In
parallel, the merging of EPs leads to multifarious Hermitian and
non-Hermitian degeneracies. It has been demonstrated that the merging of EPs
may (i) lead to the DP \cite{SLin19}; (ii) create the high-order EP~\cite%
{KDingPRX,LJinPRA18}; and (iii) form the hybrid EP~\cite{LFu,XLZhang}. The
hybrid EP has linear and square root dispersions along two orthogonal
directions and carries integer topological charge in contrast to the
ordinary EP2 that carries half-integer topological charge \cite{LeykamPRL}.

The hybrid EP can be formed by merging either two ordinary EP2s with \textit{identical} chirality [Fig.~\ref{fig1} (I) and Fig.~\ref%
{fig1} (II)] or two ordinary EP2s with \textit{opposite}
chiralities [Fig.~\ref{fig1} (III) and Fig.~\ref{fig1} (IV)]. In this paper,
we propose and investigate the \textit{later} unexplored case by
exploiting a non-Hermitian three-band lattice. The hybrid EP is created
from type III Dirac points emerging from a flat band of a three-band system
enclosed synthetic magnetic flux under the appropriate gain and loss. The
EPs can merge into different types of EPs at different system parameters.
Topological characterization employing the Berry phase and the non-Hermitian
charge/vortex that associated with the complex Riemann surface band
structure of non-Hermitian systems \cite{LeykamPRL,LFu} can not fully
distinguish all types of EPs, we employ the Berry phase and the phase
rigidity scaling exponent to overcome this difficulty. Our findings are
elaborated in a three-band system. The creating, moving, and merging of band
touching points are investigated. The DPs and various types of EPs possess
distinct topological properties, they are well distinguished from each other
under the developed topological characterization (see Table \ref{Table I})
and indicate different topological phases of the system. The topological
features of EP mergers are unveiled.

\section{The Band structure and phase diagram of the Three-band system}

\begin{figure}[tb]
\includegraphics[bb=0 0 510 320, width=8.8 cm, clip]{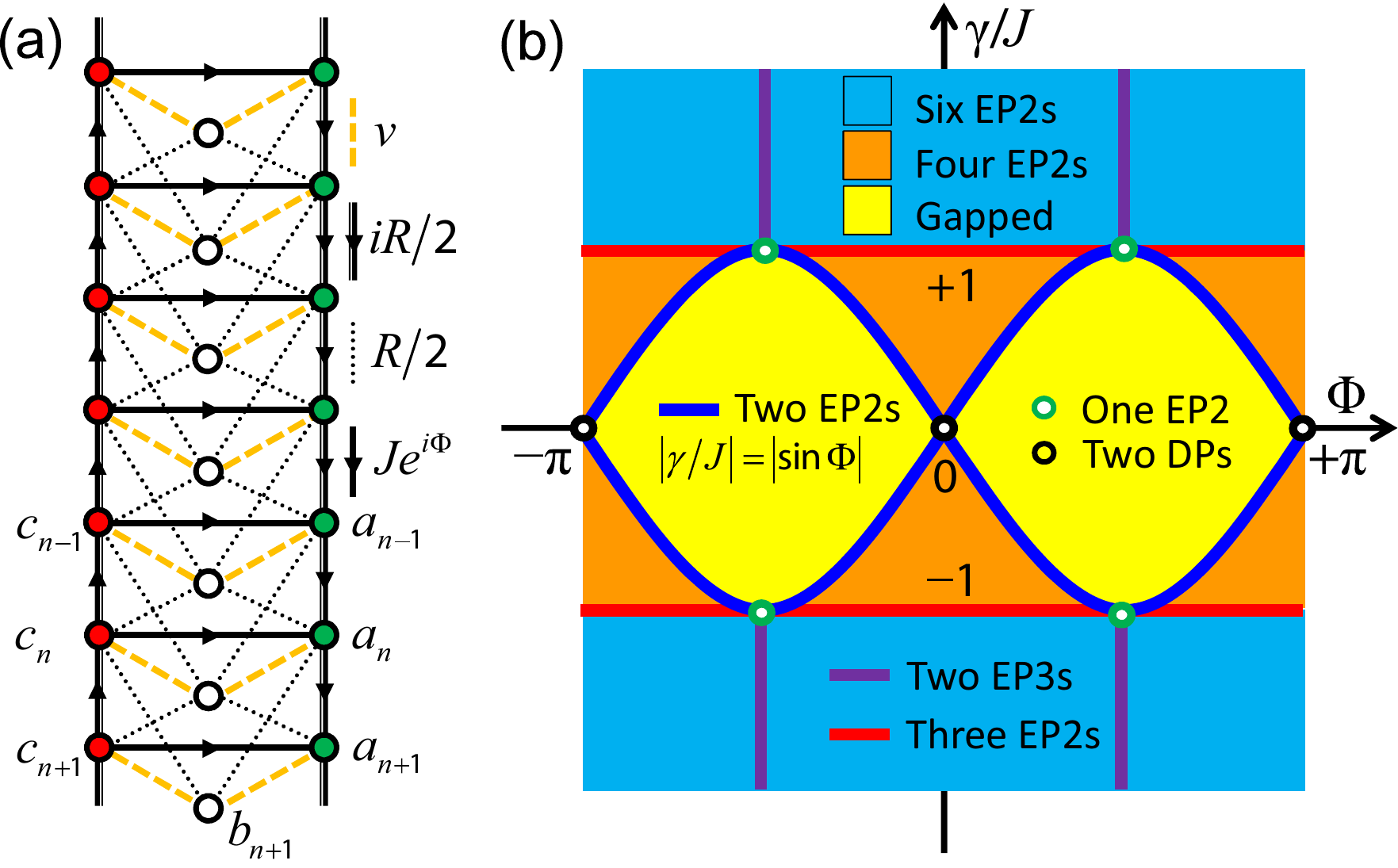}
\caption{(a) Horizontally
$\mathcal{PT}$-symmetric quasi-one-dimensional triangular lattice. $v$, $R$, $J$ are the coupling strengths among the sublattices $a$, $b$, and $c$. $\Phi$ is the Peierls phase in the nonreciprocal coupling $J e^{\pm i\Phi}$ between sublattices $a$ and $c$; the arrows
indicate the phase direction of the couplings. The green (red) site indicates the
gain (loss). (b) Phase diagram of $H$ in Eq.~(\ref{H}). The EP merging of Fig.~\ref{fig1} (III) occurs when the system parameters are chosen at the red lines and the merging of Fig.~\ref{fig1} (IV) occurs when the system parameters are chosen at the blue lines.}
\label{fig2}
\end{figure}

We consider a non-Hermitian three-band system with various
configurations of EPs in the gapless phase~\cite{SLin19}. The Hamiltonian
reads
\begin{equation}
H=\left(
\begin{array}{ccc}
h_{z}+i\gamma & h_{x} & Je^{i\Phi } \\
h_{x} & 0 & h_{y} \\
Je^{-i\Phi } & h_{y} & -h_{z}-i\gamma%
\end{array}%
\right) .  \label{H}
\end{equation}%
The investigation of $H$ in the parameter space $\left(
h_{x},h_{y},h_{z}\right) $ helps us grasp its topological properties. Since
the couplings $h_{x}$ and $h_{y}$ play the same role, we take $h_{x}=h_{y}$
without loss of generality. To be concrete, Fig.~\ref{fig2}(a) schematically
illustrates a three-band lattice. Applying the Fourier transformation, the
Bloch Hamiltonian of the three-band lattice shown in Fig.~\ref{fig2}(a) is
obtained in the form of $H$ with $h_{x}=h_{y}=v+R\cos k $, $h_{z}=R\sin k$,
where $k$ is the momentum (see Appendix A for more details). The trajectory
of $(h_{x},h_{z})$ forms a closed circle in the $h_{x}$-$h_{z}$ parameter
space, the topological features of the Bloch bands directly relate to the
topological properties of the band touching points enclosed in the
trajectory of $(h_{x},h_{z})$ \cite{TELee}. Thus, the topological properties
of the three-band lattice can be obtained by studying the band structure and
topology of $H$ in Eq.~(\ref{H}).

\begin{figure*}[tb]
\includegraphics[bb=0 0 615 290, width=17.8 cm, clip]{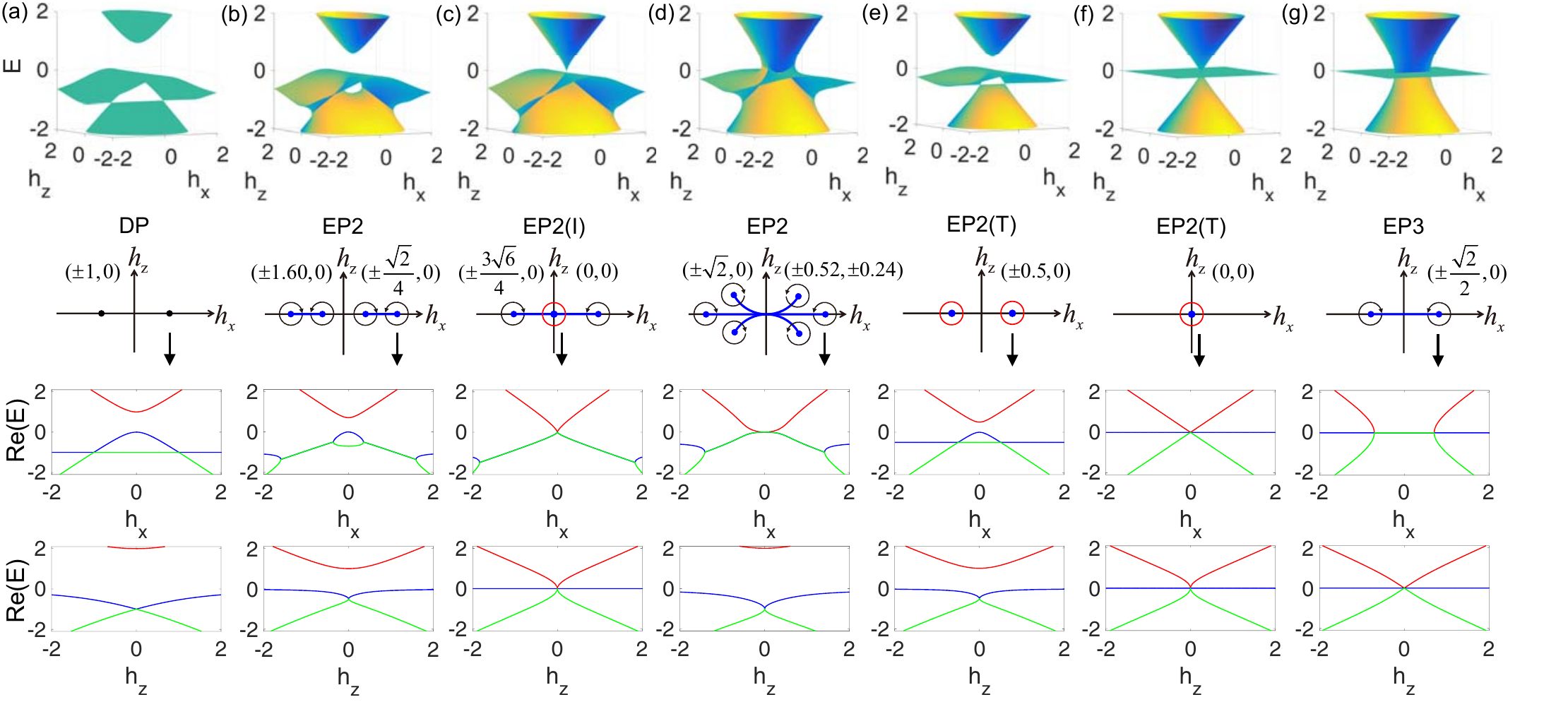}
\caption{Energy bands and EPs distribution in the parameter space for the seven typical gapless phases in the phase diagram of Fig.~%
\protect\ref{fig2}(b). In the upper panel, two axes in the horizontal plane
represent $h_{x}$ and $h_{z}$, while the vertical axis represents the real
part of eigenvalues, the color indicates the imaginary part of eigenvalues.
(a) $\Phi =0$, $\protect\gamma =0$ [black
circle
in
Fig.~\ref{fig2}(b)], (b) $\Phi =0$, $\protect\gamma =\protect%
\sqrt{2}/2$ [orange region in Fig.~\ref{fig2}(b)], (c) $\Phi
=0$, $\protect\gamma =1$ [red line in Fig.~\ref{fig2}(b)], (d) $%
\Phi =\protect\pi /3$, $\protect\gamma =\protect\sqrt{2}$ [cyan
region
in
Fig.~\ref{fig2}(b)], (e) $\Phi =\protect\pi /3$, $\protect%
\gamma =\protect\sqrt{3}/2$ [blue line in Fig.~\ref{fig2}(b)],
(f) $\Phi =\protect\pi /2$, $\protect\gamma =1$ [green
circle
in
Fig.~\ref{fig2}(b)], (g) $\Phi =\protect\pi /2$, $\protect%
\gamma =\protect\sqrt{2}$ [purple line in Fig.~\ref{fig2}(b)].
The coupling $J=1 $ in all plots. Types of EPs (blue dots) are illustrated
in the middle panels. The EPs at $\left( h_{x},h_{z}\right) $ are marked.
The $+1$ ($-1$) chirality of EPs is indicated by the counterclockwise
(clockwise) arrow, the solid blue lines are Fermi arcs connected the EPs.
The real spectra along both $h_{x}$ and $h_{z}$ directions for the
corresponding EPs are depicted in the lower panels. }
\label{fig3}
\end{figure*}

In Fig.~\ref{fig2}(a), the sublattices $a$ and $c$ are indirectly coupled
through the sublattice $b$ and are directly coupled through a nonreciprocal
coupling $Je^{i\Phi }$ \cite{LJinPRA18,LJin}. The Peierls phase factor $%
e^{\pm i\Phi }$ can be realized in various manners~\cite%
{Franz,Hafezi1,Hafezi2,Fang,ELi,TopoPhot,Zollor,Cooper,Ozawa}, which induces
effective magnetic fluxes in the triangles but not the square plaquettes of
the lattice. The synthetic magnetic fields have been experimentally realized
in coupled optical resonators \cite{Mittal1,Mittal2}. The sublattice $a $ ($%
c $) has gain (loss) and the sublattice $b$ is passive. $\mathcal{PT}$%
-symmetric systems can be investigated by employing passive system with
different losses, sticking absorption material or cutting waveguide induces
additional loss \cite{Poli,Weimann,Rechtsman}. The flat band in
non-Hermitian lattice without synthetic magnetic flux was previously
proposed through engineering the gain and loss \cite{RamezaniPRA}; besides,
it was demonstrated that proposing a spectrum that entirely constituted by
flat bands is possible with non-Hermitian couplings \cite{LeykamPRB}.
Alternatively, the flat band in the Hermitian systems maintains in the
non-Hermitian situation at the proper match of synthetic magnetic flux and
non-Hermiticity \cite{LJin19FB}. The synthetic magnetic flux provides a
useful resource to generate different types of EPs and motivates us to study
the problem of creating, moving, and merging EPs; in particularly, the
topological properties of different types of EPs.

We first consider $\gamma =0$, the lower band gap closes and the lower two
bands touch at a pair of DPs for $\Phi =0,\pi $~\cite{Phi} and $h_{z}=0$.
The DPs are two Dirac points at the peaks of two type III Dirac cones which
are the critically titled type I Dirac cones with a flat band line Fermi
surface. The type III Dirac cone associated with a line Fermi surface
differs from the type I Dirac cone associated with a point Fermi surface as
well as the type II Dirac cone associated with two cross lines Fermi surface
\cite{FLiu18,SMeng18,Amo19}. The band touching creates a flat band $%
-e^{i\Phi }J$ due to the destructive interference at the sublattice $b$. The
lower two bands have isotropic linear dispersion near the Dirac points, as
shown in the lower panel of Fig.~\ref{fig3}(a).

The DPs disappear when gain and loss are introduced ($\gamma \neq 0$), and
the band touching points become EPs. $H$ can exhibit rich band structures
featured from different types of EPs. Figure~\ref{fig2}(b) depicts the phase
diagram; it shows the number of DPs (EPs) in\ the parameter plane $h_{x}$-$%
h_{z}$ at certain fixed $\Phi $ and $\gamma /J$. The types of EPs vary in
different regions of $\gamma $-$\Phi $. The change of types of EPs indicates
the topological phase transition and the different configurations of
multiple EPs represent different topological phases of $H$ \cite{SLin19}.
Typical energy bands in the $h_{x}$-$h_{z}$ plane are exemplified in Fig.~%
\ref{fig3}.

By introducing gain and loss ($\gamma \neq 0$), the two Dirac points [Fig.~%
\ref{fig3}(a)] can split into two pairs of ordinary EP2s with opposite
chiralities [Fig.~\ref{fig3}(b)]. Hybrid EP2 is formed by merging a pair of
ordinary EP2s with opposite chiralities at an appropriate match between
non-Hermiticity and synthetic magnetic flux [Figs.~\ref{fig3}(e) and~\ref{fig3}(f)]. The distribution of different types of EPs is
indicated in the middle panel. At the band touching points, the system
parameters satisfy%
\begin{eqnarray}
\lbrack 3p^{2}-\left( 2\gamma h_{z}\right) ^{2}]\gamma h_{z} &=&0, \\
4p^{3}-27q^{2}-12p\left( 2\gamma h_{z}\right) ^{2} &=&0,
\end{eqnarray}%
where $p=2h_{x}^{2}+J^{2}-\gamma ^{2}+h_{z}^{2}$ and $q=2h_{x}^{2}J\cos \Phi
$. For $h_{x}\neq h_{y}$, $h_{x}^{2}$ is replaced by $h_{x}h_{y}$ in $p,q$.

Three types of EP2s exist in the three-band system: (i) the ordinary EP2,
which is the singularity point in the Riemann surface of square root type
and has chirality. (ii) EP2(T), which is a merger of two ordinary EP2s with
opposite chiralities. Two relevant bands touch at the EP2(T) and their real
parts are gapped in the vicinity of EP2(T), the schematic of
merging is indicated in Fig.~\ref{fig1} (IV). (iii) EP2(I), which is also
constituted by merging two ordinary EP2s with opposite chiralities. Two
relevant bands intersect, the schematic of merging is indicated in Fig.~\ref{fig1} (III). EP2(T) has anisotropic dispersions, being linear
(square root) along $h_{x}$ ($h_{z}$) while EP2(I) is has isotropic square
root dispersions along both $h_{x}$ and $h_{z}$. If $H$ has chiral symmetry,
the EPs of $H$ become EP3s.
Figure \ref{fig3} depicts the energy bands and schematically illustrates
all the seven typical EP configurations in the gapless phase. The central two ordinary EP2s in Fig.~\ref{fig3}(b) merging into one EP2(I) in Fig.~\ref{fig3}(c) is the
merging of type Fig.~\ref{fig1}(III); and the four ordinary EP2s in Fig.~\ref{fig3}(b)
merging into two hybrid EP2(T)s in Fig.~\ref{fig3}(e) is the merging of type Fig.~\ref{fig1}(IV).

The EPs may disappear when the band are gapped for $\Phi \neq 0,\pm \pi $.
This happens at weak non-Hermiticity $\left\vert \gamma /J\right\vert <|\sin
\Phi |$ [yellow region in Fig.~\ref{fig2}(b)], where three bands are gapped.
The gain and loss compress the band gaps, the lower two bands touch at the
EPs when $\left\vert \gamma /J\right\vert =|\sin \Phi |$, the appropriate
non-Hermiticity awakens the destructive interference and reproduces the flat
band. The flat band energy is altered to $-J\cos \Phi $ \cite{LJin19FB}, and
the band touching points become two EP2(T)s rather than two DPs. As the gain
and loss rates $\gamma $ increase, the band gaps vanish and all three bands
intersect when $\left\vert \gamma /J\right\vert \geqslant 1$.

\section{Topological characterization of band touching points}

The geometric (Berry) phase of energy band are relevant to the topological
features of DPs and EPs. The generalized geometric phase for the
non-Hermitian systems is defined as~\cite{Wright1988,Berry1995}
\begin{equation}
\Gamma _{n}=i\oint_{\mathrm{C}}\langle \phi _{n}\left( k\right) \left\vert
\mathrm{\nabla }_{k}|\psi _{n}\left( k\right) \right\rangle \mathrm{d}k,
\end{equation}
where $n$ is the band index. $\left\vert \psi _{n}\left( k\right)
\right\rangle $ and $\left\vert \phi _{n}\left( k\right) \right\rangle $ are
the eigenstates of Hamiltonians $H$ and $H^{\dagger }$ and form a
biorthonormal basis $\langle \phi _{n}\left( k\right) \left\vert \psi
_{n^{\prime }}\left( k\right) \right\rangle =\delta _{nn^{\prime }}$. The
integration is performed over a loop $\mathrm{C}$ in the parameter space.
For the DPs and EP2s that only relevant to two bands, the irrelevant third
band restores its original eigenvalue when the loop $\mathrm{C}$ of system
parameters encircling a band touching DP or EP for one circle; the
corresponding eigenstate accumulates a zero geometric phase; when the bands
are tangled in the presence of Hermitian or non-Hermitian band degeneracies
\cite{Mead,Lidar}, the non-Abelian Berry connections $A_{mn}=\langle \phi
_{m}\left( k\right) \left\vert \mathrm{\nabla }_{k}|\psi _{n}\left( k\right)
\right\rangle $ characterize the topological properties of energy bands \cite%
{Lidar,RYu11,Soluyanov12}. The global geometric phase $\Theta
=\sum_{n=1}^{3}\Gamma _{n}$ is a topological invariant~\cite%
{Lidar,Soluyanov,Huang}. The winding number $w=\Theta /\left( m\pi \right) $
characterizes the topology of band touching points, where $m$ is the number
of relevant bands.

Moreover, the global geometric phase is unable to distinguish the topology
of all different EPs and needs the assistance of phase rigidity
\begin{equation}
r=|\left\langle \psi _{n}^{\ast }\right. \left\vert \psi _{n}\right\rangle
/\left\langle \psi _{n}\right. \left\vert \psi _{n}\right\rangle |.
\end{equation}%
The phase rigidity $r$ describes the mixing of different states \cite{Eleuch}%
. In Hermitian system with a real matrix, the phase rigidity is $1$. When
extended to the non-Hermitian system, the defective eigenstate is
self-orthogonal and the phase rigidity at the EPs reduces to $0 $~\cite%
{Rotter09SciBull}. The phase rigidity has a scaling law in the vicinity of
EPs, $\left\vert r_{\mathrm{EP}}-r\right\vert \propto \left( \gamma _{%
\mathrm{EP}}-\gamma \right) ^{\nu }$. The phase rigidity scaling exponent $%
\nu $ characterizes the response manner of energy bands when approaching the
EPs along the parameter $\gamma $; while the geometric phase we discussed
characterizes the topological features of energy bands around EPs in the
parameter space at fixed $\gamma $.

Table~\ref{Table I} summarizes the winding number and the phase rigidity
scaling exponent along $\gamma $ for the DP and all the types of EPs. We
turn to discuss the details of topological properties of the band touching
points. The discussions as follows are organized in the order of gain and
loss increase. The creating, splitting, moving, and merging of band touching
points and their topological features are presented. We focus on the
topological features of two types of unexplored two-state coalescence:
EP2(T) and EP2(I); the details on the topological features of other band
touching points are provided in the Appendix.

\begin{table}[tbp]
\caption{Topological characterization of DP and EP (see Appendix B
and C for more details).} \label{Table I}

\begin{tabular}{c|ccccc}
\hline\hline
& DP & EP2(T) & EP2(I) & EP2 & EP3 \\ \hline
$\pm w$ & $0$ & $0$ & $0$ & $1/2$ & $1$ \\
$\nu $ & $2$ & $1$ & $1/2$ & $1/2$ & $1$ \\ \hline\hline
\end{tabular}
\end{table}

Two DPs appear at $\left( h_{x},h_{z}\right) =(\pm 1,0)$ for $\Phi =0$ at $%
\gamma =0$ as depicted in Fig.~\ref{fig3}(a), they are type III Dirac points
embedded in the flat band and appear in the region marked by the
black circles in the phase diagram of Fig.~\ref{fig2}(b). The geometric
phase for each of the two degenerate bands is $0$ when encircling either DP
for one circle; the geometric phase for the irrelevant upper band is also $0$%
. The phase rigidity scaling exponent is $\nu =2$ (see Fig.~\ref{DP} in
Appendix C).

The band gap between the lower two bands is closed in the presence of the
flat band in Fig.~\ref{fig3}(a). When the gain and loss are introduced ($%
\gamma \neq 0$), the energy bands become even closer and each DP splits into
two ordinary EP2s with opposite chiralities at the weak non-Hermiticity $%
\left\vert \gamma /J\right\vert <1$. This describes the orange
region in the phase diagram of Fig.~\ref{fig2}(b). Figure~\ref{fig3}(b)
depicts the band spectrum at the situation $\gamma /J=\sqrt{2}/2<1$. The
four ordinary EP2s are on the $h_{z}=0$ axis, and their chiralities are
opposite with respect to $h_{x}=0$ as illustrated in the middle panel. The
eigenvalues of two relevant coalescence states are $4\pi $ periodic in $k$
when encircling the ordinary EP2s, but the period of the third state is $%
2\pi $. After one circle of encircling, two coalescence relevant states
exchange and the third state restores its original eigenvalue; the global
geometric phases $\Theta $ accumulated by the three bands equals to $+\pi $ (%
$-\pi $) for the ordinary EP2 of $+1$ $(-1)$ chirality (see Fig.~\ref{QEP2}
in Appendix B), and two circles of encircling yields a $+\pi $ ($-\pi $)
geometric phase for either relevant band. The phase rigidity scaling
exponent that associated with the ordinary EP2s is equal to $\nu =1/2$ (see
Fig. \ref{EP2} in Appendix C).

\begin{figure}[tb]
\includegraphics[bb=0 0 435 280, width=8.8 cm, clip]{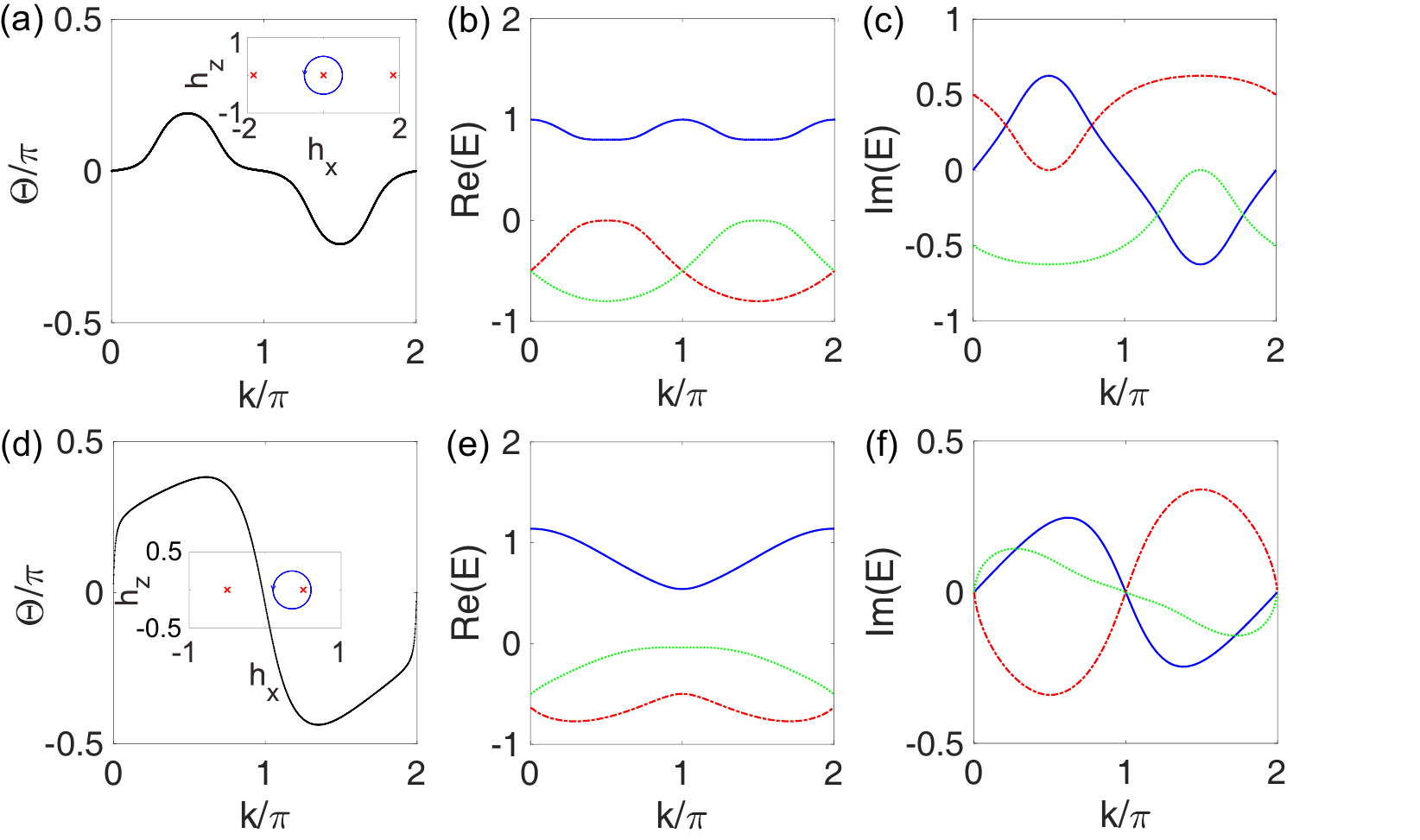}
\caption{(a-c) Encircling an EP2(I) at $J=1$, $\Phi=0$, $\gamma=1$, $v=0$, $R=1/2$ [Fig.~\ref{fig3}(c)]. (d-f) Encircling an EP2(T) at $J=1$, $\Phi=\pi/3$, $\gamma=\sqrt{3}/2$, $v=\sqrt{2}/4$, $R=1/4$ [Fig.~\ref{fig3}(e)].
The EPs are marked by the red crosses and the trajectory of encircling in the parameter space is represented by the blue circle (insets); the loop radius is $R$, centered at $(v,0)$.}
\label{fig4}
\end{figure}

\begin{figure}[tb]
\includegraphics[ bb=0 0 465 300, width=8.8 cm, clip]{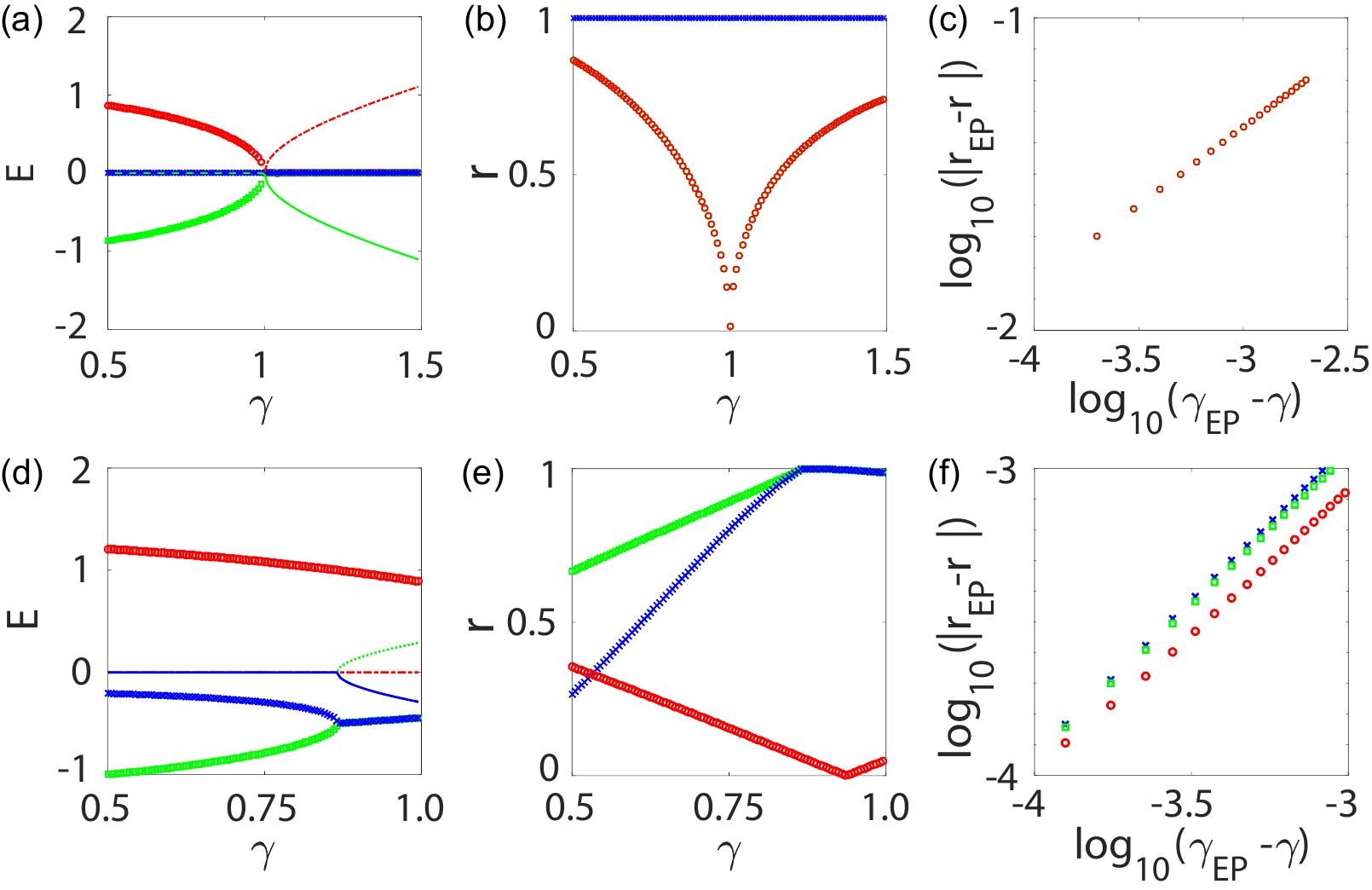}
\caption{Energy, phase rigidity, and scaling exponent (a-c) in the
vicinity of an EP2(I) in Fig.~\ref{fig3}(c) at $h_x=0$, $h_z=0$, $J=1$, $\Phi=0$ and (d-f) in the vicinity of an EP2(T) in Fig.~\ref{fig3}(e) at $h_x=1/2$, $h_z=0$, $J=1$, $\Phi=\pi/3$. The real (imaginary) part is indicated by the markers (lines).
The scaling exponent is $\nu=0.5$ for an EP2(I) and is $\nu=1.0$ for an EP2(T).}
\label{fig5}
\end{figure}

As the gain and loss rates $\gamma $ increase, the central two ordinary EP2s
become closer, but the outer two ordinary EP2s become far way. At $%
\left\vert \gamma /J\right\vert =1$, the central two ordinary EP2s with
opposite chiralities meet and merge to an EP2(I) at $\left(
h_{x},h_{z}\right) =\left( 0,0\right) $~\cite{LFu}, and the system enters the region marked by the red lines in the phase diagram of
Fig.~\ref{fig2}(b) and has three EP2s [Fig.~\ref{fig3}(c)] with identical
scaling exponent $\nu =1/2$ although they posses distinct topology [Figs.~%
\ref{fig5}(a)-\ref{fig5}(c)]. At the EP2(I), although three levels have
identical zero energy; only two levels coalesce and they degenerate with the
third level; the system is defective with one eigenstate missing. The energy
levels restore their original values after encircling the EP2(I) for one
circle; states switch twice for the lower two levels; the geometric phases
for the two lower levels are still opposite, being $\pi $ and $-\pi $,
respectively; and the winding number for the global geometric phase is $0$.
The global geometric phase for encircling the EP2(I) at $\left(
h_{x},h_{z}\right) =\left( 0,0\right) $ is depicted in Fig. \ref{fig4}(a).
The real and imaginary parts of eigen energies of $H_{k}$ for the encircling
process are depicted in Figs.~\ref{fig4}(b) and~\ref{fig4}(c). The other two
EP2s $(\pm 3\sqrt{6}/4,0)$ are ordinary EP2s. The EP2(I) in the center is
connected to the other two ordinary EP2s by two Fermi arcs \cite{HZhou}.

In the region $|\sin \Phi |<\left\vert \gamma /J\right\vert <1$, four
ordinary EPs with two opposite chiralities in pairs locate on the two sides
of $h_{z}=0$, respectively [Fig.~\ref{fig3}(b)]. Through changing $\Phi $,
each pair of EPs can merge into an EP2(T) when $\left\vert \gamma
/J\right\vert =|\sin \Phi |$ for $0<\left\vert \Phi \right\vert <\pi /2$
[Fig.~\ref{fig3}(e)]; this is indicated by the blue lines in the
phase diagram of Fig.~\ref{fig2}(b). Two valleys (peaks) in the middle
(lower) band; the apexes of valleys and peaks touch and two EP2(T)s are
formed at the appropriate match between the effective magnetic flux $\Phi $
and non-Hermiticity $\gamma $, the flat band reappears. In contrast to two
DPs, the isolated band touching points are EP2(T)s located at $%
h_{x}^{2}=J^{2}\cos ^{2}\Phi $. In the parameter space, the global geometric
phase for encircling EP2(T) $\left( h_{x},h_{z}\right) =\left( 1/2,0\right) $
in the counter clockwise direction is depicted in Fig.~\ref{fig4}(d).
Figures~\ref{fig4}(e) and~\ref{fig4}(f) are the real and imaginary parts of
eigen energies of $H_{k}$. No state switch occurs when encircling the EP2(T)
in the parameter space for one cycle and the global geometric phase is $0$;
the EP2(T) has the winding number $0$ without chirality and the scaling
exponent of phase rigidity is $\nu =1$ [Fig.~\ref{fig5}(d)-\ref{fig5}(f)].
In the region of weak non-Hermiticity $\left\vert \gamma /J\right\vert
<|\sin \Phi | $, the three bands are gapped without band touching.

For $\left\vert \gamma /J\right\vert >1$, the central EP2(I) vanishes and
splits into four ordinary EP2s with two $+1$ and two $-1$ chiralities in $%
h_{z}\neq 0$ region and six ordinary EP2s exist, provided that $H$ is not
chiral symmetric [Fig.~\ref{fig3}(d)], the phase is the cyan
region in the phase diagram of Fig.~\ref{fig2}(b). Among total six ordinary
EP2s, three of them in the region $h_{x}>0$ ($h_{x}<0$) have $+1$ $(-1)$
chirality. The upper two bands coalesce at these four ordinary EP2s ($%
h_{z}\neq 0$); and the lower two bands coalesce at the other pair of
ordinary EP2s on the $h_{z}=0$ axis. The energy bands with six ordinary EP2s
is shown in Fig.~\ref{fig3}(d), $\Phi =\pi /3$ is chosen in order to observe
all the EP2s within the region $[-2,2]$ in the parameter space.

The three-band Hamiltonian $H$ is chiral symmetric when $J=0$ or $\Phi =\pm
\pi /2$, $\mathcal{C}H\mathcal{C}^{-1}=-H$, where $\langle m\left\vert
\mathcal{C}\right\vert n\rangle =\left( -1\right) ^{m+1}\delta _{m,4-n}$. A
zero mode flat band is formed under the chiral symmetry. The upper and lower
bands in Fig.~\ref{fig3}(f) constitute a hybrid conical surface, the
projection of which on the $E$-$h_{x}$\ ($E$-$h_{z}$) plane is a conus of
square root repulsion that differs from a Dirac cone or a semi-Dirac cone~%
\cite{HuangNM,SLin2017,RMP2018}. At $\left\vert \gamma /J\right\vert <1$,
the spectrum is gapped and EP vanishes. At $\left\vert \gamma /J\right\vert
=1$, two EP2(T)s merge to a single EP2(T) at $\left( 0,0\right) $; the phase is represented by the green circles in the phase diagram of Fig.~%
\ref{fig2}(b). At $\left\vert \gamma /J\right\vert >1$, the system has one
pair of EP3s with opposite chiralities at $\left( h_{x},h_{z}\right) =(\pm
\sqrt{(\gamma ^{2}-J^{2})/2},0)$ [Fig.~\ref{fig3}(g)]; the phase
is indicated by the purple lines in the phase diagram of Fig.~\ref{fig2}(b). After encircling an EP3 for one circle, the upper and lower bands switch
and two circles is needed to restore the original eigenvalues. The global
geometric phase accumulated is $+3\pi $ $\left( -3\pi \right) $ after
encircling the EP3 of chirality $+1$ $\left( -1\right) $ for one circle (see
Fig.~\ref{QEP3} in Appendix B). The scaling exponent of the phase rigidity
close to EP3 is $\nu =1$ (see Fig.~\ref{EP3} in Appendix C).

\section{Discussion and Conclusion}

The topological characterization of band touching points applies for all types of EPs that not
limited to the DP, the ordinary EP2/EP3, and the hybrid EP2s presented in
the lattice model of Fig.~\ref{fig2}(a) as listed in Table I. For the EP2
merging of types Fig.~\ref{fig1}(I) and Fig.~\ref{fig1}(II), the geometric
phase is $\pi $ and the winding number is $\left\vert w\right\vert =1$. Different types of EPs are distinguishable from their topological features.

Hybrid EP generally presents in the non-Hermitian systems that possessing
the flat band with EPs embedded. The EP embedded in a flat band is a hybrid
EP if two-dimensional parameter space is considered \cite{LeykamPRB}.
Notably, $H$ can describe non-Hermitian Lieb lattice with additional gain
and loss $\gamma $. The synthetic magnetic flux $Je^{i\Phi }$ can be induced
by the spin-orbital coupling \cite{Franz}. The Bloch Hamiltonian of the
non-Hermitian Lieb lattice has the form of $h_{x}=2J_{x}\cos \left(
k_{x}/2\right) $, $h_{y}=2J_{y}\cos \left( k_{y}/2\right) $, $h_{z}=0$ \cite%
{LiebNJP}. Besides, hybrid EP can appear in the absence of flat band and
hybrid EP of arbitrary high order is possible to be created with asymmetric
couplings \cite{XLZhang,YXXiao}. The asymmetric coupling has connection with
the gain and loss in non-Hermitian systems. In practice, the gain and loss
associated with the effective magnetic flux equivalently induce asymmetric
coupling and nonreciprocity in the non-Hermitian systems \cite%
{LJin,XZZhang13,LJin19}.

In conclusion, we propose the hybrid EP2 through merging two ordinary EP2s
with opposite chiralities, which are created from the type III Dirac points
emerging from a flat band through introducing a proper gain and loss. The
topology of degenerate (exceptional) point is characterized by the winding
number ($w$) associated with global geometric phase and phase rigidity
scaling exponent ($\nu $). The topological properties of different EP2
mergers are unveiled, the change of topological features associated with the
merging of EPs indicates the topological phase transition. Our findings pave
the way of creating, moving, and merging EPs and are valuable for future
studies on the non-Hermitian topological phase of matter. In the future,
further investigations on the dynamical encircling of hybrid EPs \cite%
{XLZhang}, the creating, moving, and merging of high-order EPs \cite%
{SLin19,LJinPRA18}, as well as the topological edge states~\cite{LJin17}
would be of great interest.

\acknowledgments We acknowledge the support of National Natural Science
Foundation of China (Grants No.~11975128, No.~11605094, No.~11604220, and
No.~11874225). B.B.W. also acknowledges the President's Fund of the Chinese
University of Hong Kong, Shenzhen. 

\appendix*

\section*{Appendix A: Triangular lattice}

\begin{figure}[thb]
\includegraphics[bb=0 0 270 270, width=8.8 cm, clip]{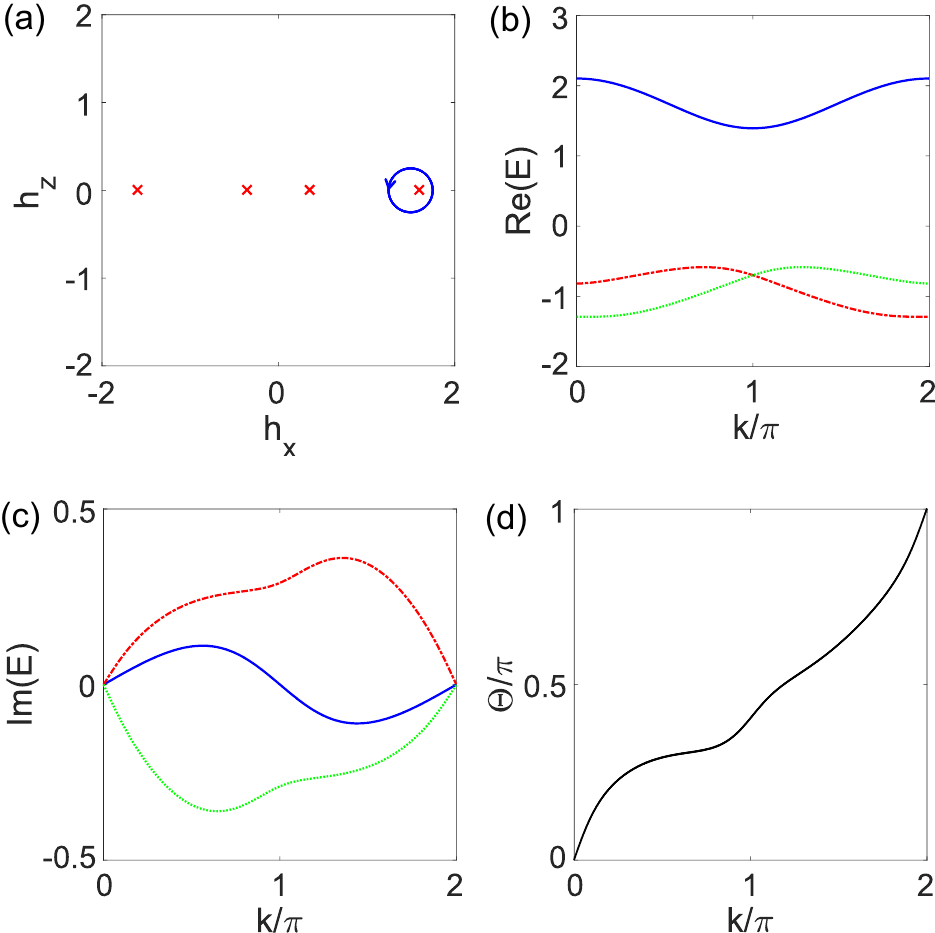}
\caption{Encircling an ordinary EP2 in Fig.~\ref{fig3}(b) with the chirality $+1$. (a) Trajectory of encircling in the parameter space is indicated by the blue circle, the EPs are marked by the red crosses. (b, c) are the real and imaginary parts of energy, and the global geometric phase is shown in (d).
The system parameters are $J=1$, $\Phi=0$, $\gamma=\sqrt{2}/2$, $v=3/2$, $R=1/4$.}
\label{QEP2}
\end{figure}

\begin{figure}[thb]
\includegraphics[bb=0 0 270 270, width=8.8 cm, clip]{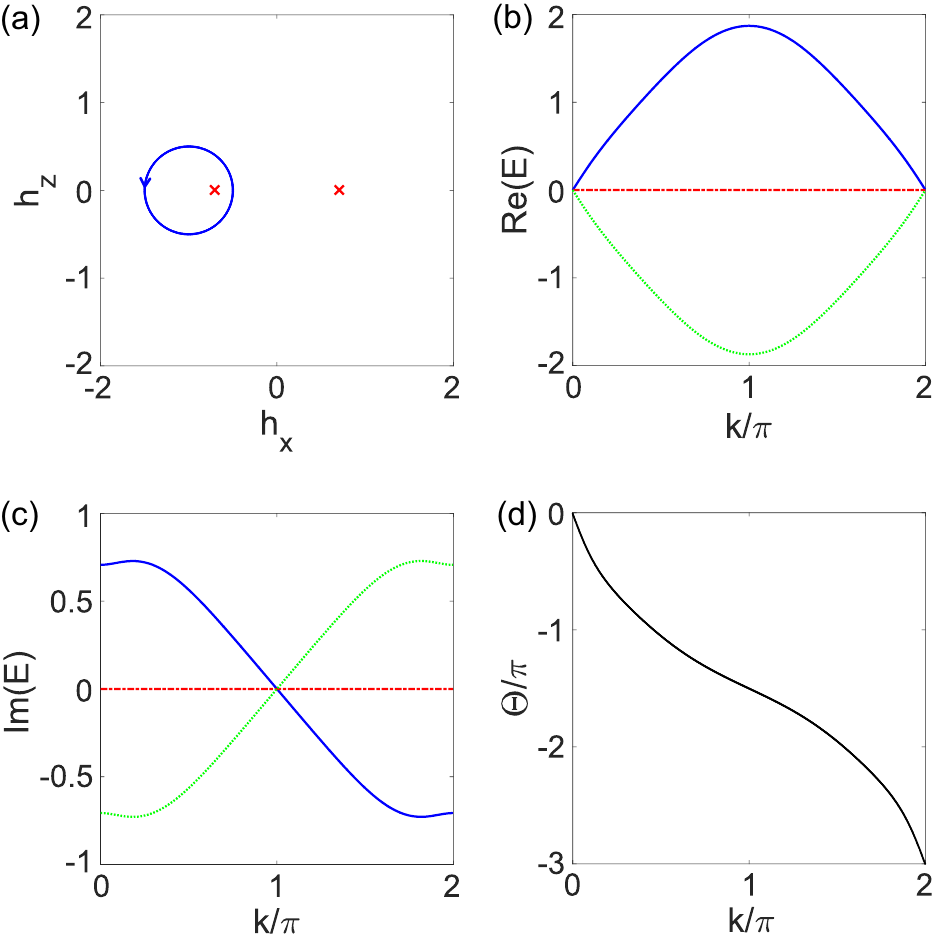}
\caption{Encircling an EP3 in Fig.~\ref{fig3}(g) with the chirality $-1$. (a) Trajectory of encircling in the parameter
space is indicated by the blue circle, the EPs are marked by the red
crosses. (b, c) are the real and imaginary parts of energy, and the global geometric
phase is shown in (d). The system parameters are $J=1$, $\Phi=\pi/2$, $\gamma=\sqrt {2}$, $v=-1$, $R=1/2$.}
\label{QEP3}
\end{figure}

\setcounter{equation}{0} \renewcommand{\theequation}{A\arabic{equation}} In
this section, we show the real space lattice Hamiltonian for the three-band
Bloch Hamiltonian $H$ in Eq.~(\ref{H}). The three-band tight-binding lattice
consists of three sublattices, the sublattice $a$ ($c$) has gain (loss) and
the sublattice $b$ is passive. The sublattices $a$ and $c$ have
nonreciprocal nearest neighbour couplings. The phase factors in the
couplings are opposite, indicated by the arrows. The sublattices $a$ and $c$
are coupled indirectly through the sublattice $b$ and directly through a
nonreciprocal coupling $Je^{i\Phi }$, which can be realized in cold atomic
gases by inducing the spin-orbital interaction, and can be realized by
optical path imbalance, dynamic modulation, and photon-phonon interaction in
optics. The triangular lattice Hamiltonian $H_{TL}$ in the real space is
given by%
\begin{eqnarray}
H_{TL} &=&\sum_{j=1}^{N}\left( -\frac{iR}{2}a_{j}^{\dagger }a_{j+1}+\frac{iR%
}{2}c_{j}^{\dagger }c_{j+1}+\mathrm{h.c.}\right)  \notag \\
&&+\left( \frac{R}{2}a_{j}^{\dagger }b_{j-1}+\frac{R}{2}a_{j}^{\dagger
}b_{j+1}+\mathrm{h.c.}\right)  \notag \\
&&+\left( \frac{R}{2}c_{j}^{\dagger }b_{j-1}+\frac{R}{2}c_{j}^{\dagger
}b_{j+1}+\mathrm{h.c.}\right)  \notag \\
&&+\left( Je^{i\Phi }a_{j}^{\dagger }c_{j}+va_{j}^{\dagger
}b_{j}+va_{j}^{\dagger }c_{j}+\mathrm{h.c.}\right)  \notag \\
&&+i\gamma a_{j}^{\dagger }a_{j}-i\gamma c_{j}^{\dagger }c_{j}.
\end{eqnarray}%
The couplings strengths are $v$, $R/2$, and $J$. The nonreciprocal couplings
$\pm iR/2$ lead to an effective magnetic flux $\pi $ enclosed in each square
plaquette. The gain and loss rates are $\gamma $. In the experimental
studies, it is not necessarily to induce the gain to balance the loss in the
investigations of $\mathcal{PT}$-symmetric lattices, using the loss only
passive systems brings the convenience \cite{Poli,Weimann,Rechtsman}. For
example, we can introduce $\left\{ 0,-i\gamma ,-2i\gamma \right\} $ instead
of $\left\{ i\gamma ,0,-i\gamma \right\} $ in the sublattices $a$, $b$, and $%
c$. By offsetting an imaginary energy $+i\gamma $ to the on-site terms $%
\left\{ 0,-i\gamma ,-2i\gamma \right\} $, we obtain the $\mathcal{PT}$%
-symmetric Hamiltonian.

Applying the Fourier transformation
\begin{equation}
a_{k}=\sum_{j=1}^{N}e^{ikj}a_{j},b_{k}=\sum_{j=1}^{N}e^{ikj}b_{j},c_{k}=%
\sum_{j=1}^{N}e^{ikj}c_{j}
\end{equation}%
where the discrete momentum is $k=2\pi n/N$ (integer $n\in \lbrack 1,N]$); $%
a_{j}$, $b_{j}$, and $c_{j}$ are the annihilation operators that satisfy the
periodical boundary condition $a_{N+1}=a_{1}$, $b_{N+1}=b_{1}$, and $%
c_{N+1}=c_{1}$. In the momentum space, the lattice Hamiltonian is expressed
as $H_{TL}=\sum_{k}H_{k}$ with $h_{x}=v+R\cos k$ and $h_{z}=R\sin k$. The
Bloch Hamiltonian $H_{k}$ in the momentum space is a $3\times 3$
non-Hermitian Hamiltonian in the form of Eq. (\ref{H}) in the main text for $%
h_{x}=h_{y}$.

\begin{figure}[t]
\includegraphics[bb=0 0 230 228, width=8.8 cm, clip]{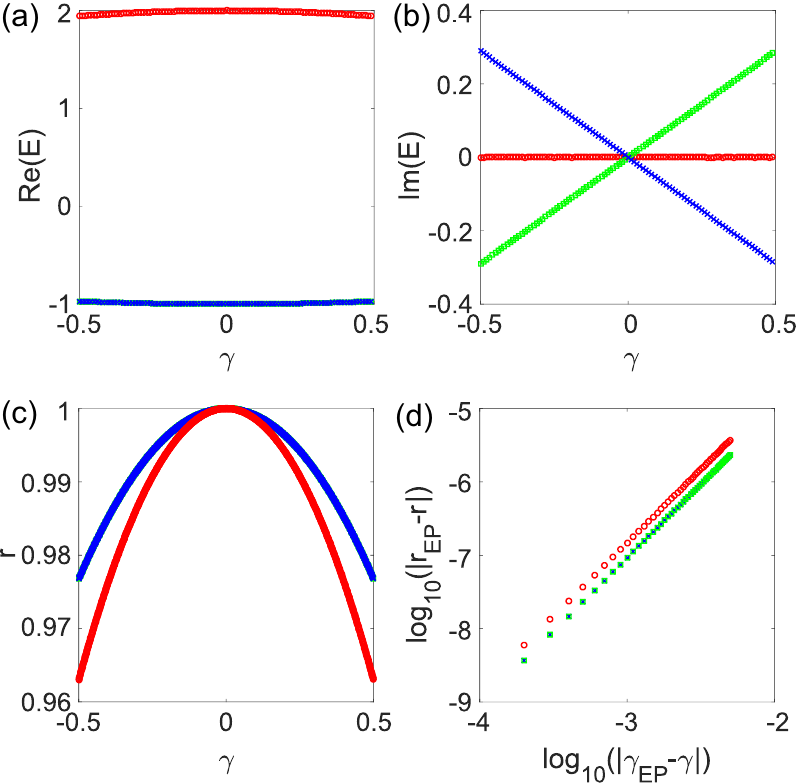}
\caption{Spectrum (a, b), phase rigidity (c), and scaling law (d) at a DP in Fig.~\ref{fig3}(a).
The system parameters are $h_x=1$, $h_z=0$, $J=1$, $\Phi=0$.
The phase rigidity scaling exponents for all the states are $\nu=2.0$.} %
\label{DP}
\end{figure}

\begin{figure}[t]
\includegraphics[bb=0 0 230 230, width=8.8 cm, clip]{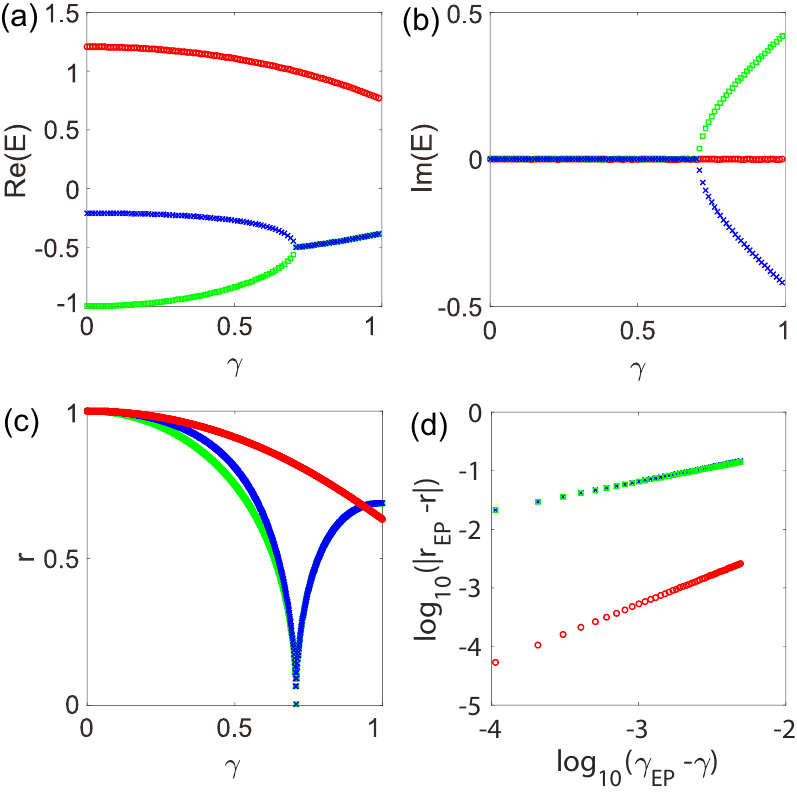}
\caption{Spectrum (a, b), phase rigidity (c), and scaling law (d) at an ordinary EP2 in Fig.~\ref{fig3}(b).
The system parameters are $h_x=\sqrt{2}/4$, $h_z=0$, $J=1$, $\Phi=0$.
The scaling exponents for the coalesced states (green square and blue cross) are $\nu=0.5$;
the scaling exponent for the third state (red circle) is $\nu=1.0$.} \label%
{EP2}
\end{figure}

\begin{figure}[t]
\includegraphics[bb=0 0 230 230, width=8.8 cm, clip]{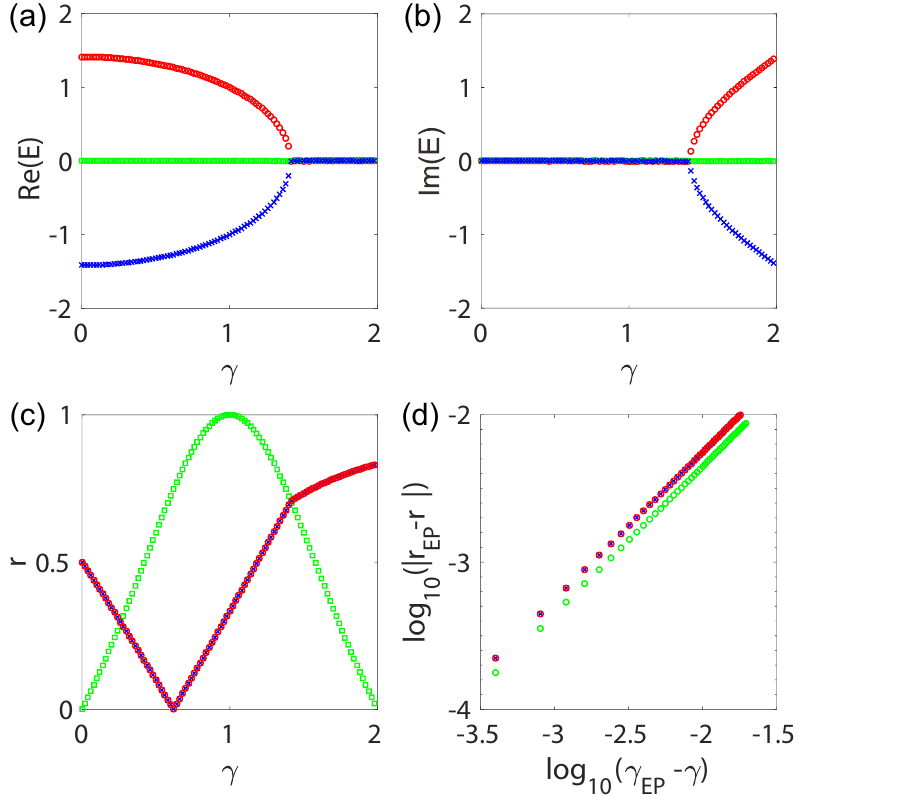}
\caption{Spectrum (a, b), phase rigidity (c), and scaling law (d) at an EP3 in Fig.~\ref{fig3}(g).
The system parameters are $h_x=\sqrt{2}/2$, $h_z=0$, $J=1$, $\Phi=\pi/2$.
The scaling exponents for all states are $\nu=1.0$.} \label{EP3}
\end{figure}
\section*{Appendix B: Geometric phase}

In this section, we show the geometric phase associated with the band
touching points, including the ordinary EP2 and EP3.

The global geometric phase for non-Hermitian system is $Q=i\sum_{n=1}^{3}%
\oint_{\mathrm{C}}\langle \phi _{n}\left( k\right) \left\vert \mathrm{d}\psi
_{n}\left( k\right) \right\rangle $ \cite{Soluyanov,Huang}. The integration
is performed over a loop $\mathrm{C}$ in the parameter space. The trajectory
of $(h_{x},h_{z}$) forms a closed circle $\mathrm{C}$ in the parameter space
of the $h_{x}$-$h_{z}$ plane.

The geometric phases for the trajectory of $(h_{x},h_{z}$) encircling an
ordinary EP2 with right chirality and encircling an EP3 with left chirality
are depicted in Fig.~\ref{QEP2} and Fig.~\ref{QEP3}, respectively. In both
cases, the coalescence associated energy levels switch after encircling the
EP once and restore their original values after encircling the EP twice. The
accumulated global geometric phase is $\pi $ for one circle of encircling
the right chiral EP2 and the accumulated global geometric phase is $-3\pi $
for encircling the left chiral EP3. The winding number is $w=\Theta /\left(
m\pi \right) $, where $m$ is the number of coalesced levels. Therefore, the
winding number for the ordinary EP2s (EP3s) is $w=\pm 1/2$ ($w=\pm 1$). The $%
+$ ($-$) sign is for the right (left) chirality.

\section*{Appendix C: Phase rigidity}

In this section, we show the band spectrum, the phase rigidity, and the
scaling exponent as the gain and loss approaching the DP, the ordinary EP2,
and the EP3, respectively.

The phase rigidity \cite{Eleuch,Rotter09SciBull} of an energy level $%
r=|\left\langle \psi _{n}^{\ast }\right. \left\vert \psi _{n}\right\rangle
/\left\langle \psi _{n}\right. \left\vert \psi _{n}\right\rangle |$ has a
scaling law in the vicinity of EPs in the form of $\left\vert r_{\mathrm{EP}%
}-r\right\vert \propto \left( \gamma _{\mathrm{EP}}-\gamma \right) ^{\nu }$.
At $\Phi =0$, the degeneracies are two-level diabolic point at $\left( \pm
1,0\right) $ in Hermitian lattice of $\gamma =0 $. In Fig.~\ref{DP}, the DPs
appear when $\gamma _{\mathrm{DP}}=0$; two energy levels become a complex
conjugation pair, the imaginary part of which changes linearly in the
vicinity of DPs. The phase rigidities for all three levels are $r_{\mathrm{DP%
}}=1.0$ and the scaling exponents are all identical $\nu =2.0$.

In Fig.~\ref{EP2}, the energy level, phase rigidity, and scaling law are
depicted for the ordinary EP2 of $\Phi =0$, $J=1$ at $(h_{x},h_{z})=(\sqrt{2}%
/4,0)$. The system has one real energy and two energy levels coalesce at the
EP2. The phase rigidities of the coalesced levels are $r_{\mathrm{EP}}=0$ at
the EP2 {$\gamma _{\mathrm{EP}}$}$=\sqrt{2}/2$, the corresponding scaling
exponent is $\nu =0.5$; the third level that not participated in the
coalescence has the scaling exponent $\nu =1.0$.

In Fig.~\ref{EP3}, the energy level, the phase rigidity, and the scaling law
are depicted for the EP3 of $J=1$, $\Phi =\pi /2$ at $(h_{x},h_{z})=(\sqrt{2}%
/2,0)$. The system has chiral symmetry, the energy levels are symmetric
about $0$ and exhibit a square root dependence on the non-Hermiticity as $%
\pm \sqrt{2-\gamma ^{2}}$. The phase rigidities of three eigenstates equal
to $r_{\mathrm{EP}}=\sqrt{2}/2$ at the EP3 $\gamma _{\mathrm{EP}}=\sqrt{2}$.
The scaling exponents are $\nu =1$.


\begin{thebibliography}{99}
\bibitem{Phi} The AB phase $\Phi $ has a period of $2\pi $. In the
discussion, we have taken $\Phi $ $(\mathrm{mod}$ $2\pi )$ and consider $%
\Phi \in \left[ -\pi ,\pi \right] $.
\end{thebibliography}
\end{document}